\documentclass[journal,onecolumn]{IEEEtran}

\usepackage{adjustbox, multirow,booktabs}

\usepackage{graphicx,caption,tikz} 
\usepackage{mathptmx, bm, amsfonts, amsmath,amsthm,breqn, listings}
\newtheorem*{lemma*}{Lemma}
\usepackage{url}
\usepackage{rotating} 
\usepackage{float}

\title{Formal Verification of Permission Voucher Protocol}
\author{
    Khan Reaz and Gerhard Wunder\\
    Cybersecurity and AI Research Group, Freie Universität Berlin\\
    Emails: \{khan.reaz, g.wunder\}@fu-berlin.de
}

\begin{document}
\maketitle

\begin{abstract}
The \emph{Permission Voucher} protocol enables privacy-preserving and secure authentication using digital ID cards in smart-city infrastructure. This paper provides a formal analysis of the protocol, outlining its design, security guarantees, and potential vulnerabilities. Using tools like the Tamarin Prover, we formally verify critical security properties, including authentication, confidentiality, integrity, and replay prevention. This work highlights the effectiveness of formal verification in ensuring robust security for modern authentication systems.
\end{abstract}

\section{Formal analysis}
Formal analysis is about building a detailed mathematical model that shows how the protocol, its users, and any attackers behave. The model includes assumptions about the communication environment, the abilities of adversaries, and the goals of the protocol. By applying logical reasoning and mathematical proof techniques, it becomes possible to verify whether the protocol satisfies these security properties or identify conditions where the protocol may fail.\\

The key objective of formal analysis is to prove that certain desired security properties hold within the model. These properties, including confidentiality, integrity, authentication, and non-repudiation, are formally defined as logical assertions that the protocol must satisfy. For instance, a confidentiality property might be expressed as, \emph{an adversary cannot deduce the contents of a message that was exchanged between two parties}. Formal methods allow for the rigorous testing of whether these properties are preserved under \emph{all} conditions.\\

Three prominent approaches within this field are \emph{process algebra, pi calculus, and symbolic models}. Each offers a unique perspective and set of tools for representing and reasoning about protocol interactions.

\subsection{Process Algebra}

Process algebra provides a framework for describing the behavior of concurrent systems, including security protocols. It represents processes as algebraic expressions and defines a set of operators for combining and manipulating these expressions. In the context of security protocols, process algebra can model the message exchanges between protocol participants, their internal state transitions, and the possible interleavings of actions.  \\

By specifying the protocol's behavior as a process algebraic expression, we can reason about its properties using algebraic laws and equivalences. This allows us to prove or disprove various security properties, such as confidentiality, authentication, and integrity, by demonstrating that the protocol's behavior satisfies certain algebraic specifications.

\subsection{Pi Calculus}

Pi calculus is a process calculus specifically designed for modeling concurrent systems with mobile processes, making it well-suited for analyzing security protocols where communication channels and data can be dynamically created and passed around. In pi calculus, processes are represented as terms that interact through channels, exchanging messages and performing internal computations. \\

The applied pi calculus extends the basic pi calculus with constructs for modeling cryptographic operations, such as encryption, decryption, and digital signatures. This allows for the representation of security protocols and their potential interactions with an adversary. Verification tools like ProVerif leverage the applied pi calculus to automatically analyze protocols and prove or disprove their security properties.

\subsection{Symbolic Models}

Symbolic models provide a more abstract representation of security protocols, focusing on the symbolic manipulation of messages and their components rather than the underlying bit-level details. They typically represent messages as terms built from symbolic constants, variables, and function symbols representing cryptographic operations.\\

The behavior of the protocol is defined by a set of rules that govern how messages are constructed, sent, received, and transformed. Adversary capabilities are also modeled symbolically, allowing for the analysis of the protocol's resilience against various attack scenarios. Tools like Tamarin Prover utilize symbolic models to explore the state space of the protocol and verify or falsify desired security properties.

\section{Verification tools}
A wide range of security protocol verification tools (and languages) have been developed, each utilising distinct methodologies and offering unique capabilities for the analysis of security protocols. We provide an overview of some of the most well-known tools in ascending order below:\\

\textbf{Isabelle/HOL}, one of the oldest  generic interactive theorem prover that requires manual input from a user in order to provide a high level of expressiveness and flexibility for formalising and verifying security protocols~\cite{isabelle1994}. \\

\textbf{Coq} is a proof assistant that enables the expression of mathematical definitions, executable algorithms, and theorems. It provides an environment for semi-interactive development of machine-checked proofs, ensuring their correctness and reliability~\cite{coq2005}.\\

\textbf{CryptoVerif} analyzes protocols in the computational model of cryptography.  It employs game-based proofs to establish security guarantees within the model~\cite{cryptoverif2007}.\\

 The \textbf{Scyther} model checker is well known for its  user-friendliness and efficiency. By automatically exploring the state space of a protocol, Scyther can rapidly identify potential attacks and generate concrete attack traces~\cite{scyther2008}. \\

 The \textbf{AVISPA} framework integrates a number of protocol analysis tools, thereby facilitating the verification process. For example Coq can be used as a back-end with AVISA's  high-level language. It supports both bounded and unbounded verification techniques~\cite{avisa2005}.\\

 \textbf{ProVerif} is based on the principles of applied pi calculus. It is able to confirm or disprove complex security properties, even when faced with the actions of a sophisticated adversary~\cite{proverif2018}.\\

The \textbf{Tamarin Prover} is an effective tool for both bounded and unbounded verification, which involves proving the absence of attacks and falsification, or the identification of attacks. It employs a symbolic model augmented with temporal logic to express and verify intricate security properties that involve temporal ordering and causality~\cite{tamarin2013}.

\begin{table}
\centering
\begin{tabular}{@{}p{2cm}p{2.5cm}p{4.5cm}p{4cm}@{}}
\toprule
\textbf{Tool} & \textbf{Type} & \textbf{Strengths} & \textbf{Shortcomings} \\ \midrule
Isabelle & Theorem Prover & High expressiveness for proofs in a wide range of domains, including cryptography. & Lacks built-in cryptographic primitives. \\

Coq & Theorem Prover & Powerful and general-purpose; can formally verify any kind of system, including cryptographic protocols & High learning curve, requiring expert knowledge of proof tactics; time-consuming for complex protocols \\

CryptoVerif & Computational Model & Computationally sound proofs, supports cryptographic primitives.& Complexity of tool usage; limited automation for large protocols. \\

Scyther & Symbolic  Model& Fast unbounded verification, falsification, and protocol analysis & Does not support computational model; limited to symbolic verification of cryptographic protocols. \\

AVISPA & Model Checker & Automated validation of Internet security protocols, efficient model-checking for various security goals & Focused mainly on Internet protocols, less flexibility with user-defined cryptographic protocols \\

ProVerif & Symbolic Model & Extensive automation, can handle complex protocols, supports secrecy, authentication, and process equivalence properties. & Poor performance  with large-scale protocols. \\

Tamarin & Symbolic Model  & Combines symbolic analysis with explicit reasoning about cryptographic primitives; supports both bounded and unbounded protocol analysis. Active community and support. & Needs to master  term/rules rewriting schema. \\

 \bottomrule
\end{tabular}
\caption{Summary  of  protocol verification tools}
\end{table}

\section{Dolev-Yao adversary model}

In 1983, Danny Dolev and Andrew Yao introduced  a foundational adversary model for analysing security protocols in the fields of computer science and cryptography~\cite{dolev1983security}. This model encapsulates \emph{three} fundamental assumptions:\\

-- The initial assumption,  \emph{\textbf{Perfect Cryptography}}, presents a somewhat idealized view of cryptographic primitives, treating them as if they were black boxes within a symbolic model. This abstraction permits the representation of cryptographic operations as function symbols within an algebra of terms, complete with defining equations. Messages can be conceptualized as algebraic terms constructed from these primitives, providing a clean and abstract method for reasoning about protocol behavior without delving into the intricacies of specific cryptographic implementations. This assumption effectively circumvents the complexities of real-world cryptographic vulnerabilities, focusing instead on the logical structure and flow of protocols.\\

    -- The second assumption, called \emph{\textbf{Unbounded Execution}}, adds another way to scale and generalize the model. This means that a protocol can be run as many times as needed by any number of agents, who can take on different roles. This is important for understanding how protocols work in the real world, where they are run many times at once on large networks with lots of people involved. It allows us to look at how different combinations of runs affect the way the protocol works.\\

    -- The third and most distinctive assumption is the concept of a \textbf{\emph{Network Adversary}}. The adversary controls the communication network and can intercept, block, replay, and spoof messages. The adversary can also make messages look like they came from anywhere and put them into the network. This model shows a worst-case scenario of network security, where communication may be vulnerable to tampering or eavesdropping. The only thing this adversary can't do is break cryptographic protections. It can only learn the contents of messages that aren't secured or for which it doesn't have the keys.


\section{Why Tamarin prover?}

We chose the Tamarin Prover for its proven capability to rigorously analyze the security of some of the real-world cryptographic protocols. In particular, Tamarin was used to propose  solution to countering the Wi-Fi \emph{KRACK} attack~\cite{KRACK-formal}.\\

One of the main advantages of Tamarin is its multiset rewriting rule-based approach. This method allows us to describe how messages are exchanged, processed, and manipulated, not only by legitimate participants but also by potential adversaries. Tamarin's symbolic modeling makes it easier to capture a wide range of protocol behaviors and test them for vulnerabilities. By simulating different attack scenarios, Tamarin can automatically prove security properties, such as authentication and confidentiality, or identify specific attacks that threaten those properties.\\

Additionally, Tamarin offers both automated and interactive modes for proof construction. This versatility is invaluable for our work. The automated mode is efficient for routine analysis, quickly verifying protocol security or identifying flaws. However, when dealing with more complex protocols or challenging verification tasks, the interactive mode allows us to manually guide the proof process, investigate potential attacks in depth, and refine the analysis. This balance between automation and manual control gives us the flexibility needed to explore both simple and sophisticated security challenges.\\

Another key advantage is Tamarin’s support for equational theories, which is crucial for protocols that rely on operations like Diffie-Hellman key exchange. This capability ensures that Tamarin can handle the advanced cryptographic techniques used in modern security protocols, making it especially useful for analyzing widely deployed systems such as Transport Layer Security (TLS) and Public Key Infrastructure (PKI).\\

Finally, Tamarin's track record in a variety of high-profile protocols such as \texttt{5G, TLS 1.2/1.3}, and \texttt{BLE} gave us the confidence to use it for the protocols we were developing.

\section{Modeling Protocols with Tamarin Prover}

To verify a security protocol using the Tamarin prover, a formal script must be constructed according to Tamarin's predefined syntax and conventions. This script is written in Tamarin's modeling language, which specifies the protocol's structure, roles, and security properties in a way that Tamarin can interpret and analyze. Each script must have the file extension \texttt{.spthy}, which stands for \emph{Security Protocol Theory}.

\subsection{Structure of a Tamarin Model}

The structure of a Tamarin model typically consists of the following key components:\\

\textbf{Declarations}: The script begins with declarations of basic components such as messages, functions, and constants. These declarations define the basic cryptographic operations (e.g., encryption, decryption, hashing) and protocol-specific constants that will be used throughout the model. For example: \verb|functions: enc/2, dec/2, hash/1| \\ 

Here, \verb|enc/2| and \verb|dec/2| represent encryption and decryption functions with two arguments, while \verb|hash/1| represents a hash function with one argument.\\

\textbf{Rules}: Protocol behavior is modeled using rules, which describe the flow of messages and actions between different entities (roles) in the system. Each rule consists of a set of premises (conditions that must be met) and consequences (actions that result if the conditions hold). Rules capture the core interactions of the protocol, such as sending and receiving messages, performing cryptographic operations, and updating the state of the system.\\

For example, a simple rule might represent a message being sent by one party and received by another:
\begin{center}
    \begin{lstlisting}[basicstyle=\ttfamily, frame=single]
    rule SendMessage:
    [ Fr(~msg), !SenderRole ]--[ msg_sent(~msg) ]->[ Out(~msg) ]
    \end{lstlisting}
\end{center}

\textbf{State Facts and Events:} Tamarin uses facts to represent both static and dynamic information. Facts can be persistent (remaining true throughout the protocol execution) or ephemeral (holding temporarily). State facts track the current state of each participant and the protocol’s progress. Events are used to record significant occurrences, such as when a message is sent or a particular security condition is met. These events can be used later to define and prove security properties.\\

\textbf{Adversary Model}: The adversary's capabilities are explicitly modeled in Tamarin scripts. By default, Tamarin assumes a Dolev-Yao adversary, meaning the attacker can intercept, modify, and inject messages into the network. The adversary model can be extended to include additional capabilities, such as controlling certain protocol participants or interacting with specific cryptographic operations.\\

Adversary actions are typically represented using rules that define how the adversary can manipulate messages or participate in the protocol. For example:

\begin{center}
    \begin{lstlisting}[basicstyle=\ttfamily, frame=single]
rule AdversaryIntercept:
    [ In(~msg) ] --[ ]-> [ Out(~msg), !Adversary ]
    \end{lstlisting}
\end{center}

\textbf{Security Properties (Lemmas)}: The final section of the script defines the security properties to be verified, expressed as lemmas. These lemmas state the conditions that should hold true for the protocol to be secure. Examples include secrecy (confidentiality of a message), authentication (verifying the identity of participants), and integrity (ensuring that messages are not tampered with).\\

To illustrate how the Tamarin Prover works, let's consider examples for symmetric encryption, asymmetric encryption, message integrity, and authentication.\\

In \emph{symmetric encryption}, both parties share a secret key. To model this, define a function \verb|enc/2| for encryption and \verb|dec/2| for decryption, along with an equation like \verb|dec(enc(m, k), k) = m|. A rule might simulate a sender encrypting a message and a receiver decrypting it. The lemma would ensure that only the intended recipient can decrypt the message, verifying \emph{confidentiality}.\\

For \emph{asymmetric encryption}, define functions \texttt{aenc/2} for encryption and \texttt{adec/2} for decryption, with \texttt{pk/1} representing the public key. The equation \texttt{adec(aenc(m, pk(k)), k) = m} models decryption using the private key. A rule would simulate a sender encrypting a message with the receiver's public key. The lemma ensures that only the receiver, with the corresponding private key, can decrypt it, verifying confidentiality and authenticity.\\

To ensure message \emph{integrity}, use a function \texttt{mac/1} for message authentication codes. A rule might model a sender generating a MAC for a message and a receiver verifying it. The lemma checks that if the MAC is valid, the message has not been altered, ensuring integrity.\\

For \emph{authentication}, define a challenge-response protocol. A rule might simulate an authenticator sending a challenge and the enrollee responding with a computed value.

\subsection{Verification Process}






 Once the protocol is modeled, Tamarin then uses its automated reasoning engine to either prove or disprove these lemmas based on the constructed model:\\

 \textbf{   Proven Lemmas:} If a lemma is successfully proven, it means the protocol satisfies the corresponding security property within the defined adversary model. This guarantees that the protocol upholds critical security attributes, such as confidentiality or authentication.\\

\textbf{    Falsified Lemmas:} If Tamarin fails to prove a lemma, it produces a counterexample trace. This trace provides a step-by-step sequence of actions or messages that demonstrate how an attacker could exploit the protocol, revealing potential vulnerabilities or weaknesses. These counterexamples are valuable in diagnosing flaws, as they allow protocol designers to understand and rectify potential attacks.

\section{Visual meaning of the dependency graphs}
The following explanation of the Tamarin dependecny graph is  extracted from~\cite{tamarin-git}.\\

If a lemma is successfully proven, it is highlighted in green; if a counterexample is found, it appears in red.  Various arrow types and colors denote different elements in the proof visualization: \\


\textbf{Black arrows}: Indicate where a produced fact is utilized by another rule.

\begin{figure}[!ht]
    \centering
    \begin{tikzpicture}
        \draw[->, thick, black] (0, 0) -- (5, 0);
    \end{tikzpicture}
\end{figure}

\textbf{Grey arrows}: Indicate the use of persistent facts or adversary knowledge by another rule.

\begin{figure}[!ht]
    \centering
    \begin{tikzpicture}
        \draw[->, thick, gray] (0, 0) -- (5, 0);
    \end{tikzpicture}
\end{figure}

\textbf{Red arrows}: Indicate when the adversary obtains a term from the network via an \texttt{Out} fact.

\begin{figure}[!ht]
    \centering
    \begin{tikzpicture}
        \draw[->, thick, red] (0, 0) -- (5, 0);
    \end{tikzpicture}
\end{figure}

\textbf{Dotted arrows}: Represent the temporal ordering or precedence of actions.

\begin{figure}[!ht]
    \centering
    \begin{tikzpicture}
        \draw[->, thick, dashed, black] (0, 0) -- (5, 0);
    \end{tikzpicture}
\end{figure}

\textbf{Green rectangular boxes}: Rule instances are depicted following a specific format. For a rule represented as \verb|LHS--[ACT]--> RHS|, the top line corresponds to the LHS (left-hand side), the middle line denotes the ACT (action), and the bottom line represents the RHS (right-hand side). Slight variations in shades of green are employed to distinguish between instances of different rules. The action line within the box specifies the timepoint variable for a given rule instance, along with the rule name and any associated set of actions.

\begin{figure}[!ht]
    \centering
    \begin{tikzpicture}
        \draw[thick, fill=green!40] (0, 0) rectangle (6, 3);

        \draw[thick] (0, 2) -- (6, 2);
        \draw[thick] (0, 1) -- (6, 1);

                \node at (3, 2.5) {\verb|Fr(~kAT)|};
        \node at (3, 1.5) {\verb|#vr.1 : Setup[]|};
        \node at (3, 0.5) {\verb|!SharedKey($A$, $T$, ~kAT)|};

    \end{tikzpicture}
    \label{fig:partitioned-box}
\end{figure}

\textbf{Grey ellipses}: Indicate actions where the adversary selects a term, such as \verb|t1|. The term \verb|!KU| represents a specific case of \verb|K|, signifying the adversary's acquisition of a term.

\begin{figure}[!ht]
    \centering
    \begin{tikzpicture}
        \draw[thick, gray] (0, 0) ellipse [x radius=3cm, y radius=0.5cm];
    \end{tikzpicture}
\end{figure}

\textbf{Black ellipses}: Denote adversary actions, such as deriving knowledge or transmitting messages.

\begin{figure}[!ht]
    \centering
    \begin{tikzpicture}
        \draw[thick] (0, 0) ellipse [x radius=3cm, y radius=.5cm];
    \end{tikzpicture}
    \label{fig:basic-ellipse}
\end{figure}

\section*{Technical Internals}

\verb|isend|: Represents an adversary sending a message to an \texttt{In} fact.

\verb|!KU|: Terms used during the construction path.

\verb|coerce|: Indicates the start of the construction path after any potential deconstruction path.

\section{Demonstrating a Replay Attack in Tamarin}

In this example, we model a scenario where an attacker can replay a message, bypassing the security mechanism. This is a common issue in protocols without proper checks for uniqueness or freshness of messages.

\begin{lstlisting}
    

    theory ReplayAttack
    begin
    
    functions: mac/2
    
    // Key registration for message integrity
    rule Register_Key:
      [ Fr(~k) ] --> [ !MacKey($A, ~k) ]
    
    // Client sends message with a MAC (no freshness check)
    rule Client_Sends_Message:
      let m = mac('message', ~k)
      in
      [ !MacKey($A, ~k) ] --> [ Out(m) ]
    
    // Server receives the message and checks MAC (no nonce or freshness)
    rule Server_Receives_Message:
      let m = mac('message', ~k)
      in
      [ In(m), !MacKey($A, ~k) ] --> [ Out('message_verified') ]
    
    // Lemma to ensure replay attack is possible
    lemma Replay_Possible:
      "Ex m #i #j.
       (In(m) @ #i & In(m) @ #j & #i < #j)"
    
    end
    
\end{lstlisting}

In the above example, the server simply checks if the received message has a valid MAC, without verifying that the message is fresh (i.e., not a replay of an old valid message).\\

Lemma \verb!Replay_Possible!  checks if the same message \texttt{(m)} can be received multiple times, indicating a replay attack.\\

Running \verb!tamarin-prover ReplayAttack.spthy --prove! will show the lemma \verb!Replay_Possible! is \textbf{falsified}, meaning Tamarin detects the possibility of a replay attack. Specifically, the output will include a trace showing that the same message was sent and verified multiple times without any checks for freshness, which indicates the success of a replay attack.\\

\begin{lemma*}[Replay Possible]
  \begin{dmath*}
    \exists \, m \, \#i \, \#j. \,
    \left(\left(\text{In}\left(m\right) \, @ \, \#i\right) 
    \land
    \left(\text{In}\left(m\right) \, @ \, \#j\right)\right) \land
    \left(\#i < \#j\right)
  \end{dmath*}

  Guarded formula characterizing all counter-examples:
  \begin{dmath*}
    \forall \, m \, \#i \, \#j. \,
      \left(\text{In}\left(m\right) \, @ \, \#i\right) 
      \land
      \left(\text{In}\left(m\right) \, @ \, \#j\right) \Rightarrow \neg\left(\#i<\#j\right)
  \end{dmath*}

\end{lemma*}

Now, we introduce a nonce (a unique value used only once) to prevent the replay attack. Each message will include a nonce, and the server will only accept fresh, non-repeated messages.

\begin{lstlisting}
theory PreventReplayAttack
begin

functions: mac/2

// Key registration for message integrity
rule Register_Key:
  [ Fr(~k) ] --> [ !MacKey($A, ~k) ]

// Client sends a message with a MAC, including a fresh nonce
rule Client_Sends_Message:
  [ Fr(~n) ]  // Generate a fresh nonce
  let m = mac('message', ~k)
  in
  [ !MacKey($A, ~k) ] 
  --> [ Out(<~n, m>), !Nonce(~n) ]

// Server checks nonce and verifies message
rule Server_Receives_Message:
  let m = mac('message', ~k)
  in
  [ In(<n, m>), !MacKey($A, ~k), not(Nonce(n)) ] 
  --> [ Out('message_verified'), !Nonce(n) ]

// Ensure replay attack is not possible due to nonce check
lemma No_Replay_Attack:
  "All n m #i #j.
   (In(<n, m>) @ #i & In(<n, m>) @ #j) ==> #i = #j"

end

\end{lstlisting}

Running \verb|tamarin-prover PreventReplayAttack.spthy --prove| will verify that no replay attacks are possible. The lemma \verb|No_Replay_Attack| will be verified, indicating that the nonce mechanism successfully prevents the same message from being accepted multiple times.

\section{Formally Verified Security Properties}

The following security properties are considered while evaluating each protocols with the Tamarin prover. \\

\textbf{Authentication} is the process of verifying the identity of a user or device before granting access to resources or services. It ensures that only authorized entities can interact with the system, reducing the risk of unauthorized access. Authentication mechanisms can range from simple password-based systems to multi-factor approaches involving biometrics, cryptographic tokens, or one-time passwords. The robustness of authentication directly impacts the overall security of the system, as weak or compromised authentication can lead to breaches.\\
    
\textbf{Mutual authentication} extends the concept of authentication by requiring both parties in a communication to verify each other’s identities. This process ensures that both the client and the server (or two devices) are genuine and trustworthy before data is exchanged. Mutual authentication is crucial in preventing impersonation attacks, where one party might otherwise be deceived by a malicious actor posing as the other. Common methods for achieving mutual authentication include TLS with client certificates or cryptographic challenge-response protocols.\\

A \textbf{Man-in-the-Middle (MitM)} attack occurs when an attacker intercepts communication between two parties without their knowledge, allowing the attacker to eavesdrop or modify the transmitted data. MitM attacks exploit the lack of secure communication channels and can lead to the leakage of sensitive information, identity theft, or the injection of malicious content. Strong encryption, mutual authentication, and secure key exchange protocols are commonly used to defend against MitM attacks by ensuring that attackers cannot decrypt or tamper with messages.\\

\textbf{Key confirmation} is a process that ensures both parties in a communication have agreed upon and are using the same cryptographic key before secure communication can proceed. This verification step guarantees that the key exchange was successful and that no third party has tampered with or intercepted the key. Key confirmation adds an extra layer of security to key exchange protocols, ensuring that data encrypted with the agreed-upon key remains confidential and secure.\\

\textbf{Uniqueness of the session} ensures that each communication session between parties is distinct and isolated from others, preventing session reuse or replay attacks. This property is typically achieved through the use of unique session identifiers or nonces, ensuring that even if an attacker intercepts session information, they cannot replicate or hijack the session. Ensuring session uniqueness is crucial in maintaining the integrity and confidentiality of data in repeated interactions between parties.\\

\textbf{Non-repudiation} ensures that a party involved in a communication or transaction cannot later deny their involvement. This property is achieved through cryptographic techniques like digital signatures, which provide verifiable proof that a specific party sent a particular message or initiated a transaction. Non-repudiation is especially important in legal and financial contexts, where accountability and proof of action are critical for trust and compliance.\\

\textbf{Confidentiality} refers to the protection of data from unauthorized access, ensuring that only intended recipients can read the transmitted information. This is typically achieved through encryption, where data is transformed into a format that is unreadable to anyone without the proper decryption key. Confidentiality is a fundamental aspect of information security, safeguarding sensitive data from eavesdropping and preserving privacy in both communication and data storage.\\

\textbf{Integrity} ensures that the data being communicated or stored remains unaltered and accurate throughout its lifecycle. It guarantees that any unauthorized changes to the data, whether intentional or accidental, can be detected. Integrity is typically enforced through cryptographic hash functions, checksums, or message authentication codes (MACs), which allow the receiver to verify that the data has not been tampered with. Maintaining data integrity is critical in preventing data corruption and ensuring trustworthiness.\\

\textbf{Replay prevention} is a security measure that protects systems against attacks where an adversary captures and retransmits valid data or authentication credentials in an attempt to gain unauthorized access. By using unique session identifiers, timestamps, or nonces in each communication, replay prevention ensures that previously captured messages cannot be reused in a fraudulent way. This mechanism is essential for maintaining the integrity of authentication and transaction processes.\\

\textbf{Key validation} ensures that a cryptographic key used in communication or encryption is valid, secure, and not compromised. Before initiating secure communication, both parties verify the authenticity and correctness of the key to confirm that it hasn’t been tampered with or substituted by an attacker. Proper key validation prevents the use of weak, expired, or maliciously altered keys, maintaining the overall security of the cryptographic system.\\

\textbf{Remark.} When presenting the protocol in the following sections, we include all the lemmas that are modeled and verified. Additionally, where appropriate, we provide dependency graphs to illustrate the relationships among these lemmas.

\section{ Formal Verification of \textbf{Permission Voucher}}

To prove the \emph{PermissionVoucher} protocol in Tamarin, we need to focus on different phases of the protocol, their respective security properties, and the associated risks. Here, we define trust relationships and channels, then we identify key security  lemmas.\\

\subsection{Trust Relationships and Channels}
\textbf{Secure Channels:}\\
\emph{Channel between ID APP and ID-CARD (NFC):}\\ Denoted as $H_{IC}$, this channel is assumed secure as it relies on physical NFC communication, involving local proximity. Confidentiality and integrity are guaranteed since the channel is established via NFC and PIN verification.\\

\noindent\emph{Channel between ID APP and OWNER (PIN Input):}\\ $H_{PIN}$ is also assumed secure due to user input of a PIN, which only the user knows. Authentication of the OWNER is provided via PIN.\\

\textbf{Public or Partially Trusted Channels:}\\
\emph{Channel between OWNER and On-premise SERVICE (UWB connection):}\\ $H_{UWB}$ is considered partially trusted, as it may provide local network access via UWB, but the security may depend on physical constraints (proximity). Messages over this channel may be intercepted, and hence the protocol must prevent impersonation and replay attacks.\\

\noindent\emph{Channel between On-premise SERVICE and ID-VERIFIER:}\\ $H_{SI}$ is partially trusted. It involves verification of data provided by the On-premise SERVICE using the ID-VERIFIER. The integrity of the transmitted data is crucial, and this channel may be vulnerable to attacks like tampering.

\subsection{Security lemmas for Permission Voucher}

\begin{lemma*}[Authentication: Ensure that only a valid ID-CARD can be unlocked by a correct PIN]
    \begin{dmath*}
      \forall \, \text{pin} \, \text{card\_data} \, \#i \, \#j. \,
      \left(\left(\text{In}\left(\text{pin\_input}\left(\text{pin}\right)\right) \, @ \, \#i\right) 
      \land
      \left(\text{In}\left(\text{nfc\_connection}\left(\text{card\_data}\right)\right) \, @ \, \#j\right)\right) \Rightarrow
      \left(\#i < \#j\right)
    \end{dmath*}
  
    Guarded formula characterizing all counter-examples:
    \begin{dmath*}
      \exists \, \text{pin} \, \text{card\_data} \, \#i \, \#j. \,
        \left(\text{In}\left(\text{pin\_input}\left(\text{pin}\right)\right) \, @ \, \#i\right) 
        \land
        \left(\text{In}\left(\text{nfc\_connection}\left(\text{card\_data}\right)\right) \, @ \, \#j\right) 
        \land
        \neg\left(\#i < \#j\right)
    \end{dmath*}
  \end{lemma*}

\begin{lemma*}[Voucher Authenticity: Ensure that only the ID APP can create a valid permission voucher]
    \begin{dmath*}
      \forall \, \text{voucherID} \, \text{data\_items} \, \text{private\_key} \, \#i \, \#j. \,
      \left(\left(\text{Out}\left(\text{sign}\left(\text{create\_voucher}\left(\text{voucherID}, \text{data\_items}\right), \text{private\_key}\right)\right) \, @ \, \#i\right) 
      \land
      \left(\text{K}\left(\text{private\_key}\right) \, @ \, \#j\right)\right) \Rightarrow
      \left(\#i = \#j\right)
    \end{dmath*}
  
    Guarded formula characterizing all counter-examples:
    \begin{dmath*}
      \exists \, \text{voucherID} \, \text{data\_items} \, \text{private\_key} \, \#i \, \#j. \,
        \left(\text{Out}\left(\text{sign}\left(\text{create\_voucher}\left(\text{voucherID}, \text{data\_items}\right), \text{private\_key}\right)\right) \, @ \, \#i\right) 
        \land
        \left(\text{K}\left(\text{private\_key}\right) \, @ \, \#j\right) 
        \land
        \neg\left(\#i = \#j\right)
    \end{dmath*}
  \end{lemma*}

 \begin{lemma*}[Voucher Integrity: Ensure that a permission voucher is not modified during transfer]
    \begin{dmath*}
      \forall \, \text{voucherID} \, \text{serviceID} \, \text{private\_key} \, \#i \, \#j. \,
      \left(\left(\text{Out}\left(\text{sign}\left(\text{redeem\_voucher}\left(\text{serviceID}, \text{voucherID}\right), \text{private\_key}\right)\right) \, @ \, \#i\right) 
      \land
      \left(\text{In}\left(\text{sign}\left(\text{redeem\_voucher}\left(\text{serviceID}, \text{voucherID}\right), \text{private\_key}\right)\right) \, @ \, \#j\right)\right) \Rightarrow
      \left(\#i = \#j\right)
    \end{dmath*}
  \end{lemma*}
  
 \begin{figure}[h]
  \centering
    \includegraphics[scale=.5]{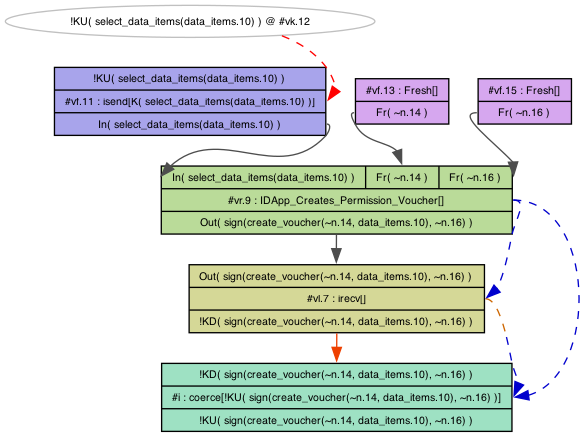}
    \caption{Tamarin dependency graph when IDApp creates PermissionVoucher}
\end{figure}

\begin{lemma*}[No Replay Nonce: Ensure that each nonce used during the process is unique and prevents replay attacks]
    \begin{dmath*}
      \forall \, \text{nonce} \, \#i \, \#j. \,
      \left(\left(\text{In}\left(\text{nonce}\left(\text{nonce}\right)\right) \, @ \, \#i\right) 
      \land
      \left(\text{In}\left(\text{nonce}\left(\text{nonce}\right)\right) \, @ \, \#j\right)\right) \Rightarrow
      \left(\#i = \#j\right)
    \end{dmath*}
  \end{lemma*}
  
\begin{lemma*}[Visitor Pass Authenticity: Ensure that a visitor pass can only be generated by the ID-VERIFIER after a valid voucher is redeemed]
    \begin{dmath*}
      \forall \, \text{passID} \, \text{voucherID} \, \text{private\_key} \, \text{verifier\_key} \, \#i \, \#j. \,
      \left(\left(\text{In}\left(\text{sign}\left(\text{redeem\_voucher}\left(\text{serviceID}, \text{voucherID}\right), \text{private\_key}\right)\right)
      \, @ \, \#i\right) \land
      \left(\text{In}\left(\text{sign}
      \left(\text{generate\_visitor\_pass}\left(\text{passID}\right), \text{verifier\_key}\right)\right)
      \, @ \, \#j\right)\right) \Rightarrow \left(\#i < \#j\right)
    \end{dmath*}
  
    Guarded formula characterizing all counter-examples:
    \begin{dmath*}
      \exists \, \text{passID} \, \text{voucherID} \, \text{private\_key} \, \text{verifier\_key} \, \#i \, \#j. \,
        \left(\text{In}\left(\text{sign}\left(\text{redeem\_voucher}\left(\text{serviceID}, \text{voucherID}\right), \text{private\_key}\right)\right) \, @ \, \#i\right) 
        \land
        \left(\text{In}\left(\text{sign}\left(\text{generate\_visitor\_pass}\left(\text{passID}\right), \text{verifier\_key}\right)\right) 
        \, @ \, \#j\right) \land
        \neg\left(\#i < \#j\right)
    \end{dmath*}
  \end{lemma*}
  
\begin{sidewaysfigure}
  \centering
    \includegraphics[scale=.5]{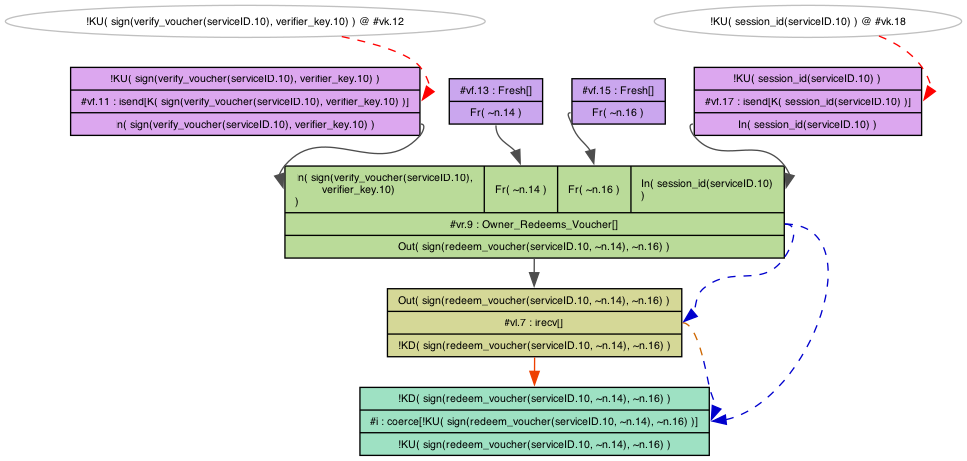}
    \caption{Tamarin dependency graph when OWNER redeems voucher}
\end{sidewaysfigure}

\begin{lemma*}[Data Items Confidentiality: Ensure that selected data items remain confidential unless explicitly shared by the OWNER]
    \begin{dmath*}
      \forall \, \text{data\_items} \, \#i \, \#j. \,
      \left(\left(\text{Out}\left(\text{select\_data\_items}\left(\text{data\_items}\right)\right) \, @ \, \#i\right) 
      \land
      \left(\text{K}\left(\text{data\_items}\right) \, @ \, \#j\right)\right) \Rightarrow
      \left(\#i = \#j\right)
    \end{dmath*}
  
    Guarded formula characterizing all counter-examples:
    \begin{dmath*}
      \exists \, \text{data\_items} \, \#i \, \#j. \,
        \left(\text{Out}\left(\text{select\_data\_items}\left(\text{data\_items}\right)\right) \, @ \, \#i\right) 
        \land
        \left(\text{K}\left(\text{data\_items}\right) \, @ \, \#j\right) 
        \land
        \neg\left(\#i = \#j\right)
    \end{dmath*}
  \end{lemma*}

  \begin{sidewaysfigure}
    \centering
    \includegraphics[scale=.5]{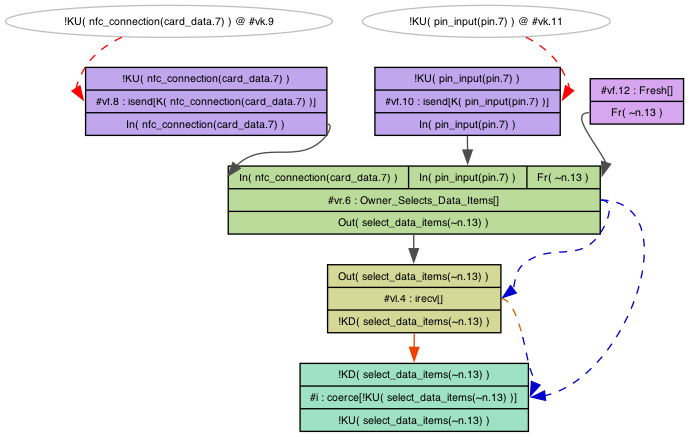}
    \caption{Tamarin dependency graph when OWNER selects card data items}
\end{sidewaysfigure}

\begin{figure}[!ht]
    \centering
    \includegraphics[scale=.5]{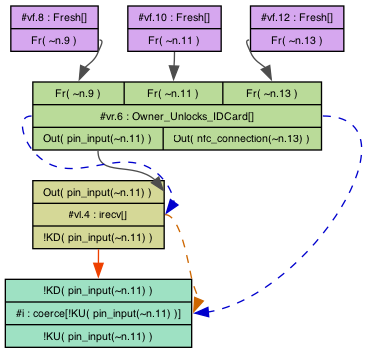}
    \caption{Tamarin dependency graph when OWNER unlocks IDCard}
\end{figure}

\begin{lemma*}[Mutual Authentication]
    \begin{dmath*}
      \forall \, \text{serviceID} \, \text{verifier\_key} \, \#i \, \#j. \,
      \left(\left(\text{In}\left(\text{sign}\left(\text{verify\_voucher}\left(\text{serviceID}\right), \text{verifier\_key}\right)\right) \, @ \, \#i\right) 
      \land
      \left(\text{In}\left(\text{session\_id}\left(\text{serviceID}\right)\right) \, @ \, \#j\right)\right) \Rightarrow
      \left(\#i < \#j\right)
    \end{dmath*}
  
    Guarded formula characterizing all counter-examples:
    \begin{dmath*}
      \exists \, \text{serviceID} \, \text{verifier\_key} \, \#i \, \#j. \,
        \left(\text{In}\left(\text{sign}\left(\text{verify\_voucher}\left(\text{serviceID}\right), \text{verifier\_key}\right)\right) \, @ \, \#i\right) 
        \land
        \left(\text{In}\left(\text{session\_id}\left(\text{serviceID}\right)\right) \, @ \, \#j\right) 
        \land
        \neg\left(\#i < \#j\right)
    \end{dmath*}
  \end{lemma*}

\begin{figure}
  \centering
    \includegraphics[scale=.5]{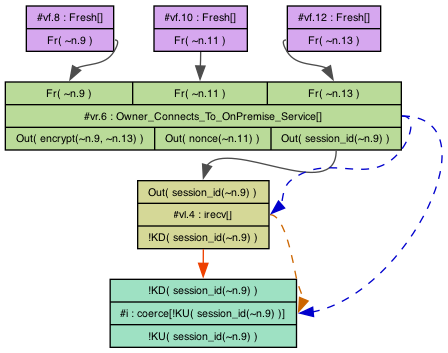}
    \caption{Tamarin dependency graph when OWNER connects to OnPremise Service}
\end{figure}

\subsection{Tamarin output for Permission Voucher}

Table~\ref{table:permissionvoucher-Tamarin} presents a summary of the output from the Tamarin prover for each security lemma.
\begin{table}[ht]
    \centering
    \begin{tabular}{@{}l l@{}}
        \hline
        \textbf{Analyzed:} & permissionvoucher.spthy \\
        \textbf{Processing Time:} & 0.25s \\
        \hline
        \textbf{Authentication (all-traces):} & verified (4 steps) \\
        \textbf{Voucher\_Authenticity (all-traces):} & verified (4 steps) \\
        \textbf{Visitor\_Pass\_Authenticity (all-traces):} & verified (4 steps) \\
        \textbf{Data\_Items\_Confidentiality (all-traces):} & verified (4 steps) \\
        \textbf{Mutual\_Authentication (all-traces):} & verified (4 steps) \\
        \hline
    \end{tabular}
    \caption{Tamarin output summary for Permission Voucher}
    \label{table:permissionvoucher-Tamarin}
\end{table}

\subsection{Identified weaknesses}

The ID-VERIFIER is trusted to validate the voucher and generate the visitor pass, but it could be vulnerable if compromised. Adding additional security checks, such as mutual authentication between the On-premise SERVICE and ID-VERIFIER, would ensure both entities are verified before any sensitive information is exchanged.\\

The current model doesn't account for potential attacks involving compromised parties (e.g., On-premise SERVICE acting maliciously). Adding role-specific lemmas to analyze potential insider threats could strengthen the model. For instance, add a lemma to verify that the ID-VERIFIER is the only entity capable of authenticating the voucher, preventing other entities from replaying or impersonating the verification process. Ensure that the On-premise SERVICE is properly registered and bound to the verification process. This could prevent adversaries from pretending to be the On-premise SERVICE.\\

There is no current check for the expiry of permission vouchers. Implementing a mechanism for verifying the time validity of a voucher (e.g., using timestamps) would prevent expired vouchers from being reused by adversaries. A \verb|Voucher_Expiry| lemma could ensure that any voucher presented for validation is within its valid time frame.\\

The model currently assumes that all keys (e.g., \verb|private_key, verifier_key|) are securely managed and known to the respective parties. Introducing key distribution and management rules would add another layer of security to the protocol, reducing the risk of keys being compromised.

\section{Known Limitations }

We note  the following  key limitations associated with using the Tamarin prover for verifying security protocols, as well as some current development challenges within the framework.\\

\textbf{Proof Strategy.} Tamarin operates within a symbolic model, whereby cryptographic operations are abstracted into symbolic terms, rather than being represented by fully detailed cryptographic functions. Consequently, Tamarin is only capable of detecting attacks that adhere to the aforementioned symbolic representation and a predefined set of equations, collectively known as subterm-convergent theories. Although it is capable of supporting some built-in functions, its capacity to handle more complex or non-standard cryptographic theories remains limited. Furthermore, due to the intrinsic complexity of the problems it addresses, Tamarin may not always complete its analysis, particularly for protocols with extensive or intricate state spaces. In such instances, it may be necessary to intervene manually by proving individual lemmas or diagnosing the reason for the failure of the trace analysis to terminate.\\

\textbf{Trace Mode vs. Observational Equivalence Mode.}  
    Tamarin employs two principal modes of analysis: \emph{trace mode} and \emph{observational equivalence mode}. Trace mode is typically regarded as a reliable and comprehensive approach, capable of identifying all potential attack traces within the confines of the symbolic model. However, the observational equivalence mode, which aims to demonstrate that two distinct executions of a protocol are indistinguishable to an adversary, remains a work in progress. Although it is a reliable method to some extent, it may fail to identify certain attacks due to the necessity for precise alignment between the rules. Furthermore, observational equivalence mode does not yet fully support all protocol features, particularly restrictions or complex message structures, which can result in incomplete analysis.\\
    
\textbf{Scalability and Performance.}  
    The complexity of formal verification represents a significant challenge for Tamarin, particularly in the context of large or state-heavy protocols, where the scalability of its analysis may be limited. In the case of protocols comprising a multitude of roles, message exchanges, or state transitions, the time required for Tamarin to complete its analysis may increase exponentially. Such limitations may result in performance bottlenecks and, in some instances, impede the prover's ability to reach a conclusion within a reasonable timeframe.\\

\bibliography{security-models}
\end{document}